**Adaptive control of the**
**singularly perturbed chaotic systems**
**based on the scale time estimation**
**by keeping chaotic property**

**Mozhgan Mombeini, Ali Khaki Sedigh, Mohammad Ali Nekoui**

Science and Research Branch, Islamic Azad University, Hesarak, Punak, Tehran, Iran
E-mail: M.Mombeini80@gmail.com, Sedigh@kntu.ac.ir, Manekoui@eetd.kntu.ac.ir

**Abstract**
In this paper, a new approach to the problem of stabilizing a chaotic system is presented. In this regard, stabilization is done by sustaining chaotic properties of the system. Sustaining the chaotic properties has been mentioned to be of importance in some areas such as biological systems.
The problem of stabilizing a chaotic singularly perturbed system will be addressed and a solution will be proposed based on the OGY (Ott, Grebogi and Yorke) methodology. For the OGY control, Poincare section of the system is defined on its slow manifold. The multi-time scale property of the singularly perturbed system is exploited to control the Poincare map with the slow scale time. Slow scale time is adaptively estimated using a parameter estimation technique. Control with slow time scale circumvents the need to observe the states. With this strategy, the system remains chaotic and chaos identification is possible with online calculation of lyapunov exponents.
Using this strategy on ecological system improves their control in three aspects. First that for ecological systems sustaining the dynamical property is important to survival of them. Second it removes the necessity of insertion of control action in each sample time. And third it introduces the sufficient time for census.

**1. Introduction**
Nonlinear singularly perturbed models are known by dependence of the system properties on the perturbation parameter [5]. Multi time scale characteristic is an important property of this class of nonlinear systems. For this class of systems a two-stage procedure for design composite controller is presented in [9]. On the other hand, chaotic behavior is an important characteristic of a class of nonlinear systems. Many researchers have shown interest in the analysis and control of the chaotic systems. Among the proposed approaches is the control of the Poincare Map (the OGY-Method) [8].
 In this paper, the OGY method is applied to the singularly perturbed chaotic systems. The proposed control strategy exploits the chaotic property of the system and a discrete system model on the Poincare map is defined. This Poincare map lies on the slow manifold of the system. It is shown that by using the two time scale property of the system, an OGY control with slow time scale on the slow manifold of system, could be defined.

This strategy of control results in keeping chaotic property of the system and then online identification of chaos with calculation of lyapunov exponents is possible. An adaptive parameter estimation technique is used to estimate perturbation parameter and the slow time scale of the singularly perturbed system. Population models are examples of systems where sustaining the dynamical property of controlled systems is important for survival of them. Chaotic model of food chains were initially found in [2,4]. Recently, chaotic impulsive differential equations are used in biological control [6,11-13]. Multi time scale approach was first used in [7] for food chain models. Method is implemented on a prey-predator type of population model.

The paper is organized as follows. In section 2 the slow-fast manifold separation based on the slow and fast states for singularly perturbed systems is introduced.
In section 3 an adaptive estimation technique for the estimation of perturbation parameter is proposed.
In section 4 chaotic property of the system is exploited and the OGY control is implemented for the stabilizing problem. Then singularly perturbed property is exploited and slow manifold of the system is selected as the Poincare section. Then a new control based on the slow scale time estimation is introduced.
Section 5 presents the results of employing the proposed method on the ecological prey-predator system.

**2. Problem Formulations**
In this paper, chaotic singularly perturbed systems of the following form are considered,

$$\varepsilon \dot{x} = f(x, y)$$
$$\dot{y} = g(x, y) \text{ (1)}$$

Where, $x \in R$ , $y \in R^{n-1}$ and $\varepsilon$ is a small parameter. $f : R^n \rightarrow R$ , $g : R^n \rightarrow R^{n-1}$ are both smooth functions and the system is chaotic.

The slow manifold of (1) is defined with

$$0 = f(x, y)$$
$$\dot{y} = g(x, y) \quad (2)$$

This $S$ manifold $S : \{f = 0\}$ is smooth and results in separation of time scales as $x$ the fast, and $y$ as the slow variable. It is easily seen that,

$$\frac{\dot{y}}{\dot{x}} = \frac{g(x, y)}{\varepsilon f(x, y)} \xrightarrow{\varepsilon \langle 0} \frac{\dot{y}}{\dot{x}} \propto \frac{1}{\varepsilon} \quad (3)$$

By taking

$$\tau = \frac{t}{\varepsilon} \quad (4)$$

the second scale time of system is $T = \dfrac{1}{\varepsilon}$ .

### 3. Adaptive Estimation of $\varepsilon$

In [10] an estimation method for constant terms using the least-squares approach is proposed. Here the method is used here for $\varepsilon$ estimation. The estimated $\varepsilon$ is found to minimize the total prediction error as

$$J = \int_0^t e^2(r) dr$$

Where the prediction error $e(t)$ is defined as

$$e(t) = \hat{\varepsilon} g(x, y)\dot{x} - f(x, y)\dot{y}$$

This total error minimization can average out the effects of measurement noise. The resulting estimation is [4]

$$\hat{\varepsilon} = \frac{\int_0^t \dot{x} g(x, y)\dot{y} f(x, y) dr}{\int_0^t \dot{x}^2 g(x, y)^2 dr} \quad (5)$$

To reduce the size of manipulations we defined $w$ window , then (5) changes to (6)

$$\hat{\varepsilon} = \frac{\int_{t-w}^t \dot{x} g(x, y)\dot{y} f(x, y) dr}{\int_{t-w}^t \dot{x}^2 g(x, y)^2 dr} \quad (6)$$

### 4. OGY Control Based On Second Time Scale Estimation

In this part a new control strategy is proposed such that controlled system remains chaotic. This strategy exploits OGY method to design control and then uses two time scale property of the system to improve the designed control such that system remains chaotic.

**4.1 Fast Direction Properties**

Consider the chaotic singularly perturbed system (1). As $\varepsilon$ is a small parameter, an approximation of the fastest eigenvalue of Jacobian matrix (7) is $\dfrac{1}{\varepsilon} \times \dfrac{\partial g}{\partial z}$ . Since the chaotic systems are dissipative and the absolute value of the sum of negative exponents is bigger than the sum of the positive Lyapunov exponents, this big value is almost negative. It means that calculation of Lyapanov exponents in fast direction is not necessary in chaos identification. And for system with one perturbation term the fast direction is a stable direction.

$$ J = \begin{bmatrix} \dfrac{1}{\varepsilon} \times \dfrac{\partial f}{\partial x} & \cdots & \cdots & \dfrac{1}{\varepsilon} \times \dfrac{\partial f}{\partial y_{n-1}} \\ \dfrac{\partial g_1}{\partial x} & \dfrac{\partial g_1}{\partial y_1} & \cdots & \dfrac{\partial g_1}{\partial y_{n-1}} \\ \vdots & \cdots & \cdots & \vdots \\ \dfrac{\partial g_{n-1}}{\partial x} & \dfrac{\partial g_{n-1}}{\partial y_1} & \cdots & \dfrac{\partial g_{n-1}}{\partial y_{n-1}} \end{bmatrix} \qquad (7) $$

**4.2 OGY Control On The Slow Manifold**

In the OGY control design, a manifold is defined such that the discrete model of the system will be obtained by the intersections of this manifold with system trajectories. Then, the control of this discrete model on this manifold will result in the control of the system. It is obvious that the manifold approach will result in a more accurate control of system if it contains all unstable modes of the system. Stable modes lead the system dynamics toward a desired point. One of the modes should be eliminated to have a manifold with unit co-dimension. Eliminating the fastest stable mode and letting it to be free leads to a more accurate control (compared to the elimination of other modes).

Considering Jacobian matrix (7) Where $x_{eq}$ is the value on the fixed point of the system while the fixed point is calculated as:

$$ f(x_{eq}, y_{eq}) = 0 $$
$$ g(x_{eq}, y_{eq}) = 0 \qquad (8) $$

Then, the discrete model will be:

$$ y_{k+1} = p(y_k, u_k), x = x_{eq} \qquad (9) $$

The OGY control, $u_k$ proposed with the following strategy on slow manifold will be:

$$ u_k = \begin{cases} K(y_k) & if\left(\left|y_k - y_{eq}\right| \leq \Delta \right. \\ \quad 0, & otherwise \end{cases} \qquad (10) $$

Where $\Delta$ is the dead zone in traditional OGY method and $K(y_k)$ is a control for slow states of discrete model (9) designed with a suitable method for example, proportional feedback.

**4.3 New OGY Control Based On Slow Time Scale Estimation**

In the OGY method, control of the Poincare map is equivalent to the control of the chaotic system (1). According to two time scale property of the system, to control this Poincare section on the slow manifold, it is sufficient to control it with the slow scale time, because the states on the slow manifold have slower motions than total dynamical system. Hence, control is designed by following strategy: Control starts with the OGY control and as soon as the first section with the Poincare map is detected, system could be controlled with the slow scale time.

Suppose that T is the estimation of slow scale time of the system. And $k_0$ is the time of first section or first pulse, the system can be controlled by inserting control action (10) only in the following instants;

$$k = k_0 + nT, n = 0,1,2,\ldots$$

Where, according to (4) T is

$$T = \left[\frac{1}{\varepsilon}\right] (11)$$

and $[\ldots]$ is a bracket symbol.

an approximation of the fastest mode will be $\dfrac{1}{\varepsilon} \times \dfrac{\partial g}{\partial z}$ . Then the accurate manifold is defined by the equation; $0 = f(x, y)$ which is the slow manifold of the system. In other expression with this strategy, the Poincare section in this problem lies on the slow manifold (2). For stabilizing problem in fixed point Poincare section becomes:

$$S = \left\{ y : x = x_{eq} \right\} \qquad (12)$$

The main idea of this new control method is keeping the system on its chaotic state without resisting to be settled down in the desired rejoin.

The control strategy can be summarized as; when the chaotic system states enter the dead zone, by insertion of the control pulse, the states settle more in the neighborhood of the slow manifold. Afterward system works in open loop and remains by its dynamic in the slow manifold. This slow manifold contains all unstable modes that are also all slow. If unstable modes try to abduct trajectory from the desired point, it needs a time. Since smaller $\varepsilon$ result in bigger fast stable modes then the time that system remains on slow manifold increases too, an approximation of this time is slow time scale of the system.

During this time after the application of the control pulse, states of the system remain in the neighborhood of the desired trajectory. Then, after this time before the exit from the desired region, the loop is closed again and insertion of an enough effective control pulse returns the trajectories closer in the slow manifold. Then system becomes open loop again and so on.

**Result 1:** control of Poincare map and control of system (1) are equivalent. With $T$ period the system is controlled. Then all needed information to control the Poincare map exist at $k = k_0 + nT$ . Then $T$ is the sufficient census time (sufficient period of observation) for system (1).

**Result 2:** By this method between the pulses system is open loop. Then system remains chaotic and online calculation of the Lyapunov exponents result in positive maximum lyapaunov exponent. By defining $\Gamma$ as

$$\Gamma = u(\lambda_{max}) = \left\{ \begin{array}{l} 0, \lambda_{max} \leq 0 \\ 1, \lambda_{max} \rangle 0 \end{array} \right\} (13)$$

When $\Gamma = 1$ the system identified as chaotic and control rule (10) could be inserted adaptively.

**4.4 Algorithm**

According to the above discussions a new algorithm to adaptive OGY control for the one term singularly perturbed systems is proposed as follows:

**Step 0:** By the slow manifold (12) construct the Poincare map (9) and design $K(y_k)$ appropriately to control this discrete model.

**Step 1:** At the first time $t$ that condition (10) is satisfied, insert the impulse control $u_k$. $pulses = 1$

**Step 2:** During $t$ to $t+w$ estimate $\varepsilon$ using (5) to estimate the slow scale time ($T$) with (11).

$census = census + w$

**Step 3:** do no act till $t = t+T$ If condition (10) is satisfied insert control $u_k$.

$pulses = pulses + 1$

**Step 4:** back to 2.

**5. Simulation Results**

In this section, the planned algorithm of section 4 is implemented on the Rosenzweig–MacArthur model. The system is model of food chains of prey-predator type. Chaotic property of the system in some range of parameters is proved in [1,3]. This model includes three states: a prey ($x$), a predator ($y$) and a top-predator ($z$), with the following equations:

$$\varepsilon \frac{dx}{dt} = x(1 - x - \frac{y}{\beta_1 + x})$$

$$\frac{dy}{dt} = y(\frac{x}{\beta_2 + x} - \delta_1 - \frac{z}{\beta_2 + y}) \qquad (14)$$

$$\frac{dz}{dt} = \xi z(\frac{y}{\beta_2 + y} - \delta_2)$$

Where

$$\beta_1 = 0.3, \beta_2 = 0.1, \delta_1 = 0.1, \delta_2 = 0.62, \xi = 0.3$$

Problem of stabilizing equilibrium point of saddle type is addressed. The Poincare section is on the following slow manifold

$$S = \left\{ (y,z) : x = x_{eq} \right\}$$

Extinction of species is not desired. While, the equilibrium point with positive and nonzero terms are desired (of biological significance). Desired fixed point is $(0.8593, 0.1632, 0.1678)$.

To design OGY and new method of control, Poincare section is linearized, and proportional feedback is used to control it. For efficiency of the method, close loop poles selected enough faster than the fastest stable pole.

Figures (1) shows the result of stabilizing with OGY control and new method. It indicates that the stabilizing with new method converges to results of OGY method. New method has lower accuracy only in the early times. But the numbers of inserted pulses decreased considerably in comparison to OGY method of control (approximately proportional to $\left[\dfrac{1}{\varepsilon}\right]$ ).

Figures (2) shows the lyapunov exponents under new method. It indicate that maximum lyapunov exponent is positive and the condition $\Gamma = 1$ for insertion the control rule (10) is satisfied and positive lyapunov exponent are in slow directions.

Figure (3) shows slow variations of states in neighbourhoud of slow manifold and effect of control pulses on the staes under new method. It indicates that in interval between the pulses, the states have slow variations.

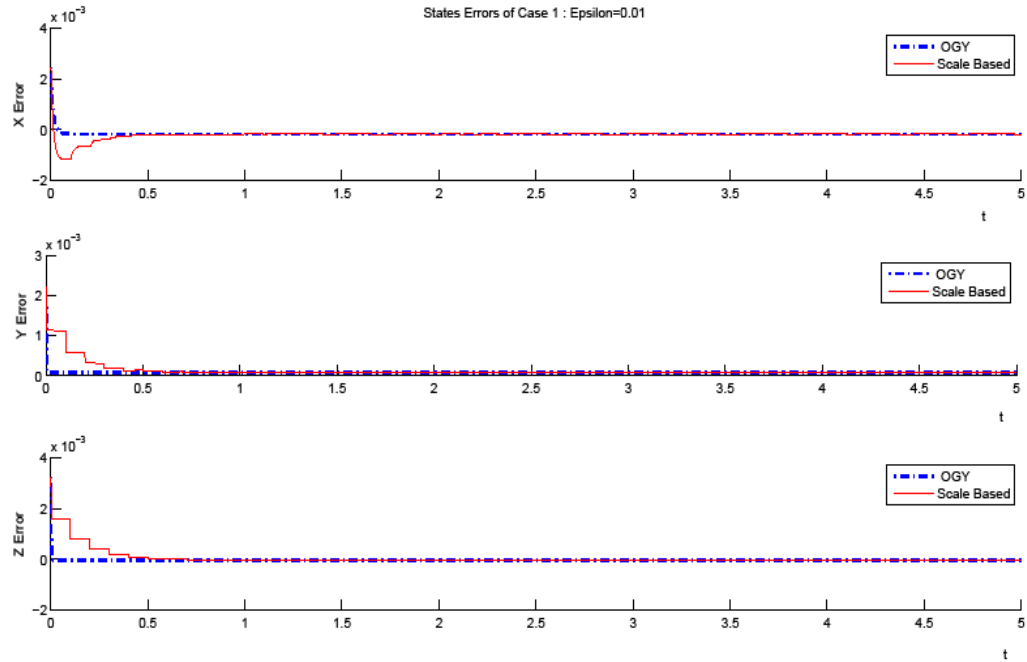

Figure (1) Comparison of the states errors by OGY control and new method (for $\varepsilon = 0.01$).

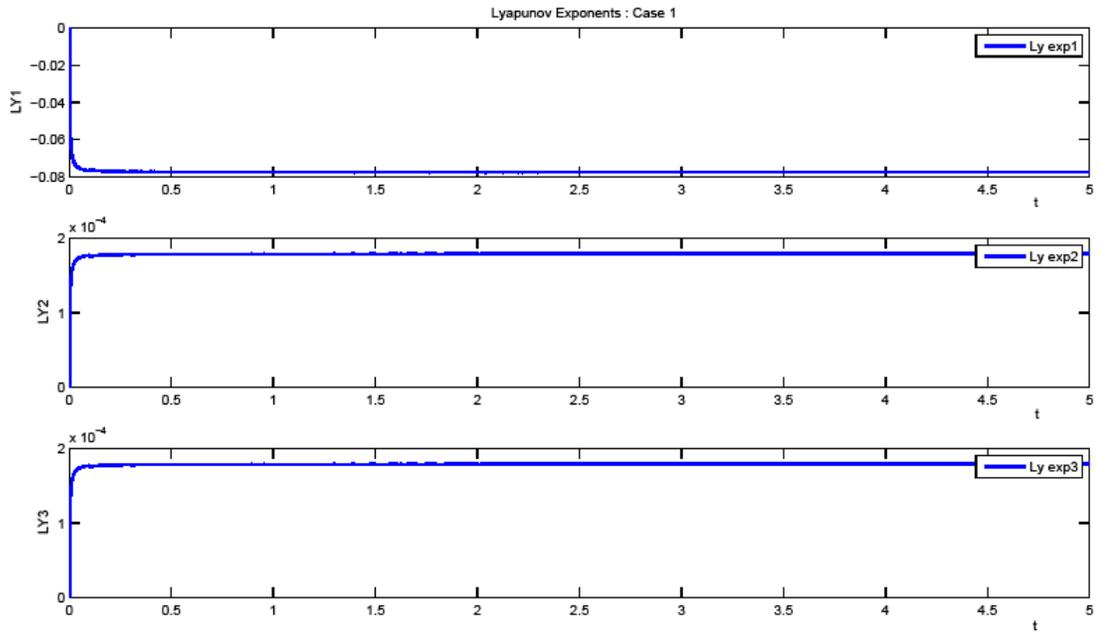

Figure (2) Lyapunov exponents of the controlled system by new method (for $\varepsilon = 0.01$).

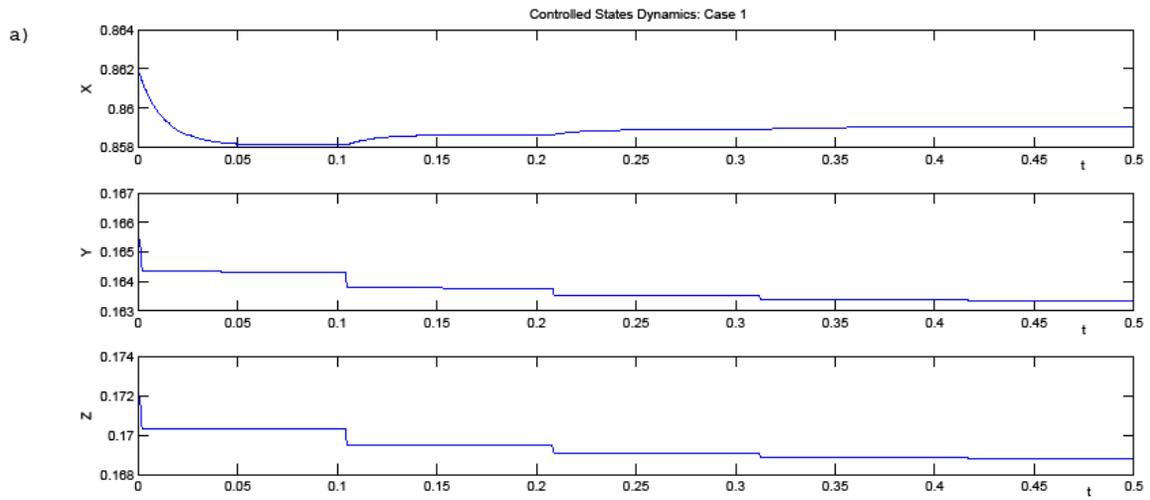

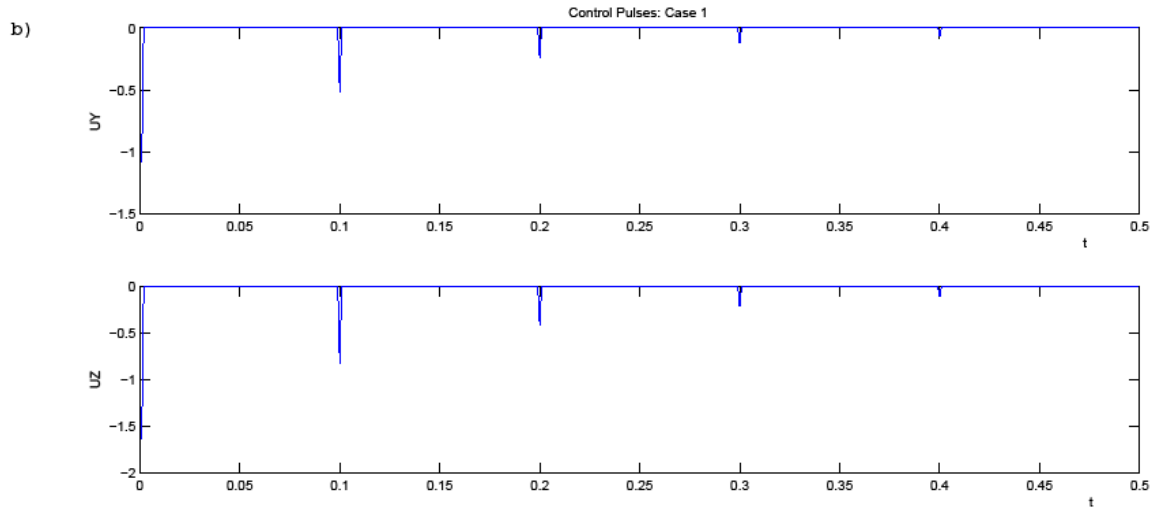

Figure (3)   a) variation of controlled states by new method in neighbourhood of slow manifold, b) inserted control pulses (for $\varepsilon = 0.01$).

This variations are such slow that dynamic of this open loop situation remains in the neighborhood of the desired trajectory. Each time insertion of the control pulses approaches systems more to the desired trajectory.

## 6. Conclusions

The simulation results on ecological model satisfying the efficiency of the new method. In proposed OGY control on slow manifold, instead of trying to drive the system trajectory to a stable rejoin, system is guided to a dynamical unstable slow manifold. Since that instability is slow, by applying the control pulses in proper times, states of system remains in neighborhood of the desired point. One of the advantages of this control strategy is that it removes necessity of observation of states for all samples. This is very important for situations that census has high expenditure (for example in biological populations) or for situation that dispatch of control action has higher expenditure (for example in pesticide). Also, maximum Lyapunov exponents remain positive and then it is useful for online adaptive identification of chaotic property of the system.